\documentclass[final]{aipproc}
\layoutstyle{8x11double}

\SetInternalRegister\hbadness{8000} 

\def\Epk{E_{\rm pk}}

\def\E0{E_{\rm 0}}
\def\P0{\varphi_{\rm 0}}
\def\Ep0{E_{\rm p0}}

\def\N0{N_{\rm 0}}
\def\t0{\t_{\rm 0}}

\def\ab{\alpha-\beta}
\def\a2{\alpha+2}
\def\b2{\beta+2}

\newcommand{\ltsima} {$\; \buildrel < \over \sim \;$}
\newcommand{\gtsima} {$\; \buildrel > \over \sim \;$}
\newcommand{\lta} {\lower.5ex\hbox{\ltsima}}
\newcommand{\gta} {\lower.5ex\hbox{\gtsima}}

\begin{document}

\title
      [Determining Bolometric Corrections for BATSE Burst Observations]
      {Determining Bolometric Corrections for BATSE Burst Observations}
 
\classification{}

\keywords{Gamma-ray bursts}

\author{Luis Borgonovo}{
  address={Stockholm Observatory, SCFAB, SE-106 91 Stockholm, Sweden}
}

\iftrue
\author{Felix Ryde}{
  address={Center for Space Science and Astrophysics, Stanford
University, Stanford, CA 94305, USA}
 }

\author{Miguel de Val Borro}{
  address={Stockholm Observatory, SCFAB, SE-106 91 Stockholm, Sweden}
}

\author{Roland Svensson}{
  address={Stockholm Observatory, SCFAB, SE-106 91 Stockholm, Sweden}
}

\fi

\copyrightyear  {2001}

\begin{abstract}
We compare the energy and count fluxes obtained by integrating over
the finite bandwidth of BATSE with a measure proportional to the
bolometric energy flux, the $\varphi$-measure, introduced by Borgonovo
\& Ryde.  We do this on a sample of 74 bright, long, and smooth
pulses from 55 GRBs.  The correction factors show a fairly constant behavior
over the whole sample, when the signal-to-noise-ratio is high
enough. We present the averaged spectral bolometric correction for the
sample, which can be used to correct flux data. 

\end{abstract}

\date{\today}

\maketitle

\section{Introduction}

The gamma-ray burst (GRB) spectrum is often peaked in the $E F_{\rm E}$
representation (where $F_E$ is the energy flux) at
some energy $\Epk$ in the $\gamma$-ray band \cite{band93}. The limited
spectral coverage of the used detector might affect the assigned
measure of the total energy-integrated flux.  
To obtain a true bolometric flux a correction is needed. 
This is most important for the photon flux and less affects the energy flux
as the energy spectrum is often peaked within the detector energy
window.

To circumvent this issue Borgonovo \& Ryde \cite{BR01} proposed to use
the value of $E F_{\rm E}$ at $\Epk$ as a representation of the energy
flux, since it is proportional to the bolometric flux. We denote this
quantity by {$\varphi$}. It was shown that the power-law
hardness-intensity correlations (HIC), that was studied for pulse
decays, were better if the $\varphi$ value was used instead of
integrating the energy flux over the BATSE band. This could indicate
that the latter method suffers from effects which are not part of the
correlation.  The $\varphi$-method thus provides an efficient way of
studying detailed features in the observed HICs.

This measure is limited to the 
cases where the peak actually exists within the studied band
$[E_{\rm min} , E_{\rm max}]$. Representing the spectrum by the
Band et~al.\ model \cite{band93}, this means that the low and high
energy power law indices must have $\alpha > -2$ and $\beta < -2$.
In the most common case where $E_{\rm max} >    (\alpha - \beta)
E_{\rm pk} /  (\alpha + 2)$, the  proportionality between ${\varphi}$
and F=$\int F_{\rm E} dE$  becomes

\begin{eqnarray}
\frac{F}{\varphi} & \equiv & \lefteqn{k(\alpha,\beta,y_{\rm min},y_{\rm max}) =
\frac{e ^{(\a2)}} {(\a2)^{\a2}} } \nonumber  \\
& & \times \left[ \vphantom{\frac{(\ab)^{\ab} y_{\rm max}^{\b2}-
(\ab)^{\a2} }{ (\beta+2) e^{\ab}}} \Gamma (\a2) \{ P(\a2,\ab)-P(\a2,
y_{\rm min}) \} \right.
\nonumber  \\
& & + \left. \frac{(\ab)^{\ab} y_{\rm max}^{\b2}-  (\ab)^{\a2} }{
(\beta+2) e^{\ab}} \right] ,
\label{k}
\end{eqnarray}

\noindent where $y_{\rm min}=(\alpha +2)E_{\rm min}/E_{\rm pk} $,
and $y_{\rm max}=(\alpha +2)E_{\rm max}/E_{\rm pk}$.
$\Gamma(\alpha +2)$ and $P(\alpha +2, y)$ are the gamma function
and the incomplete gamma function, respectively (see, e.g., Press
et al.\ \cite{Press}).  In the case that $\alpha$ and $\beta$ do
not have a strong dependence on  $E_{\rm pk}$,
 the only $E_{\rm pk}$ dependence in
$k(\alpha, \beta, y_{\rm min}, y_{\rm max})$ is in $y_{\rm min}$
and $y_{\rm max}$. In particular, when the integration is chosen
over the whole energy range from $0$ to $\infty$, there will be no
dependence at all. The {$\varphi$}-value can be interpreted as
the integral over the whole energy range or, at    least, over a
range for which $y_{\rm min} << 1$ and $y_{\rm max} >>   1$, and
should, under some circumstances, be a better representation of
the bolometric flux for the study of the hardness-intensity correlation.

\section*{Bolometric Correction to the Energy Flux}

For this work, we use a sample of 74 pulses taken from 55 GRBs,
observed by the Burst and Transient Source Experiment (BATSE) on the
{\it Compton Gamma-Ray Observatory}. The burst data have a time
resolution of multiples of 64 ms and the count flux is obtained by
adding up the four energy channels from the Large Area Detectors
(LADs) of BATSE.  We search within bright bursts (with peak fluxes
larger than 2 photons ${\rm s^{-1} cm^{-2}}$) for cases containing
long, smooth pulse structures with a general ``fast rise-slow
decay''. The light curve for each burst in the sample was rebinned to
achieve a signal-to-noise-ratio ($S/N$) of at least 30. All selected
pulses have at least 4 time-bins at the given $S/N$.  The
spectrum for each time-bin was fitted with the Band et al.\
\cite{band93} model, allowing a deconvolution to find the energy
spectrum and the peak energy.  The energy spectrum was integrated over
the energy band of the BATSE detector ($20-2000$ keV), for the
strongest illuminated LAD, to find the instantaneous energy flux.


\begin{figure}
  \resizebox{18pc}{!}{\includegraphics{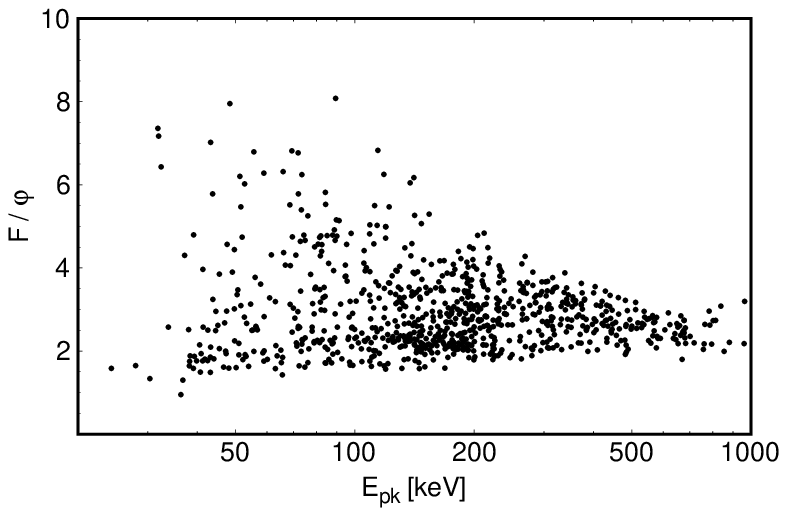}} \caption{Ratio
  $F/\varphi$ versus $\Epk$ for 846 time-resolved spectra from 55
  GRBs. The integration of the energy flux $F$ was made over the range
  $20-2000$ keV.
  }
  \label{fig1}
\end{figure}

In Figure~\ref{fig1}, the spectral bolometric correction for 846
time-resolved spectra from the 55 GRBs is shown. Average values can be
used to correct the energy flux data when $\Epk = \Epk (t)$ is known.
The data show a greater dispersion for smaller $\Epk$. This is most
likely a bias introduced by the fact that most of the spectra
correspond to the long decay phase of pulses, during which the
intensity is positively correlated with the hardness.  Therefore, the
signal-to-noise ratio is usually much higher at the peak of
the pulses when the spectra are {\it harder}, i.e., for higher $\Epk$,
and decreases to the chosen limit ($S/N=30$) as the pulse decays.

\section*{Bolometric Correction to the Count Observations}

We now redo the above analysis on the count flux $F_{\rm c}$
instead. The count data have not been deconvolved and for proper physical
interpretation the effective correction must be estimated and
understood.

\begin{figure}
  \resizebox{18pc}{!}{\includegraphics{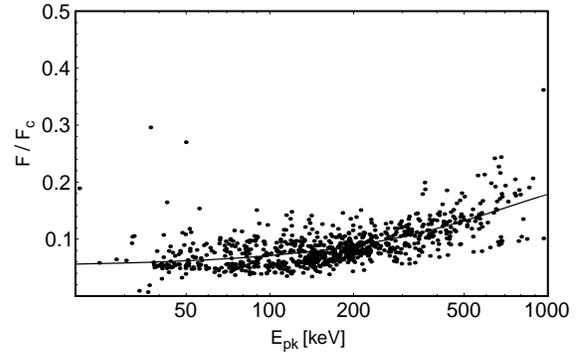}} \caption{Ratio
  of the instantaneous energy flux  to the count flux  versus the
  corresponding $\Epk$ (in keV), for the same sample of time-resolved
  spectra as in Fig.~\ref{fig1}. A best-fit second-order polynomial
  $0.05235+1.924\times10^{-4}\Epk -6.624\times10^{-8}\Epk^{2}$ is also
  shown. The dispersion parameter is $0.1516$.}  \label{fig2}
\end{figure}

Figure~\ref{fig2} shows the ratio of the instantaneous energy flux $F$
(in keV ${\rm s^{-1}} {\rm cm^{-2}}$) to the corresponding count flux
$F_{\rm c}$ (in counts ${\rm s^{-1}}$) for the same time-resolved
spectra as in Figure 1. The data were fitted with a second-order
polynomial which was found to be $0.05235+1.924\times10^{-4}\Epk
-6.624\times10^{-8}\Epk^{2}$ (with $\Epk$ given in keV).  The
dispersion was measured as the mean of the squared ratios between the
fit residuals and the fit expected values. The $F/F_{\rm c}$ fit has a
dispersion of 0.15.

\begin{figure}
  \resizebox{18pc}{!}{\includegraphics{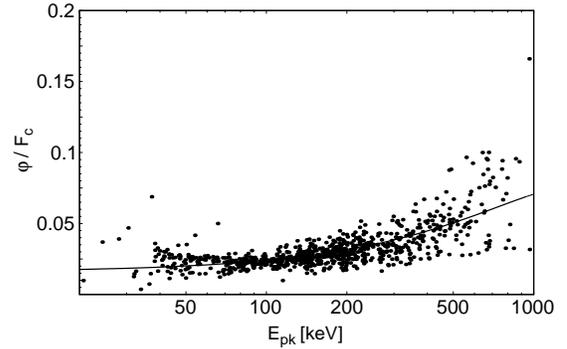}}
  \caption{Ratio of the $\varphi$-measure to the count flux $F_{\rm c}$ 
  versus $\Epk$. A second-order polynomial fit to the data $0.01597+
  8.320\times10^{-5}\Epk-2.829\times10^{-8}\Epk^2$ is also shown.
  The dispersion parameter is $0.07605$.}
  \label{fig3}
\end{figure}

In Figure~\ref{fig3}, the ratio of the $\varphi$-measure (in keV ${\rm
s^{-1}} {\rm cm^{-2}}$) to the count flux (in counts ${\rm s^{-1}}$)
over the same sample of spectra is shown. Again the data were fitted
with a second-order polynomial given by $0.01597+
8.320\times10^{-5}\Epk-2.829\times10^{-8}\Epk^2$ (with $\Epk$ in
keV). This adjusts the data with higher accuracy for small $\Epk$
values. The calculated dispersion is 0.08, significantly smaller than
in the $F/F_{\rm c}$ case, shown in Figure~\ref{fig2}.

\section*{Discussion}

To interpret the BATSE spectral data, one must have in mind that due to the
GRB spectra being broad compared to the observed energy window, one must
apply a bolometric correction factor which is unknown. To circumvent
this problem we propose the use of the $\varphi$ measure that is
also proportional to the bolometric flux. 

The flux $F$ is found from integrating the deconvolved spectrum.  This
spectrum is model-dependent as the deconvolution is based on the model
spectrum. This could introduce additional scatter into the
determination.  A second problem, when one aims at studying single
pulses in the light curve, arises from the fact that the observed
spectra may contain contributions from other pulses which even could
be unresolved. Furthermore, additional soft components \cite{Preece}
could also affect the measured flux value and thus weakening the
correlations. This makes the $\varphi$-method better to use.

The $F/\varphi$ ratio shows an approximately constant behavior over
the studied sample, with a larger dispersion for smaller $\Epk$.  This
fact can be explained as a bias introduced by the HIC correlation.

We also analyze the correction for the count flux. The
$\varphi/F_{\rm c}$ ratio is shown to have a substantially smaller
dispersion than the $F/F_{\rm c}$ case. Therefore, the effective
correction for the count flux may be better estimated.

\begin{theacknowledgments}
This research made use of data obtained through the HEASARC Online
Service provided by NASA/GSFC. We are also grateful to GROSSC, STINT,
NOTSA, and the A. E. W. Smitts Fund at Stockholm University for
support.
\end{theacknowledgments}



\begin{thebibliography}{}

\bibitem{band93} Band, D., et~al., {\it ApJ} {\bf 413}, 281 (1993).
\bibitem[Borgonovo and Ryde(2001)]{BR01}Borgonovo, L., and Ryde, F., {\it ApJ} {\bf 548}, 770 (2001).
\bibitem{Press}Press, W. H., Teukolsky, S. A., Vetterling, W. T., and
    Flannery, B. P., {\it Numerical Recipes in Fortran} 2nd Ed.,
    Cambridge Univ. Press, Cambridge, 1992.
\bibitem{Preece}Preece, R. D., Briggs, M. S., Pendleton, G. N., et al.,
    {\it ApJ} {\bf 473}, 310 (1996).




\end{thebibliography}

\end{document}